# Mechanism-based Tuning of Room-temperature Ferromagnetism in Mn-doped β-Ga$_2$O$_3$ by Annealing Atmosphere

Xu Dai, Xi Zhang, Gang Xiang*

College of Physics, Sichuan University, Chengdu 610000, China.

*Corresponding email: gxiang@scu.edu.cn

**ABSTRACT**

Mn-doped β-Ga$_2$O$_3$ (GMO) films with room-temperature ferromagnetism (RTFM) are synthesized by polymer-assisted deposition and the effects of annealing atmosphere (air or pure O$_2$ gas) on their structures and physical properties are investigated. The characterizations show that the concentrations of vacancy defects and Mn dopants in various valence states and lattice constants of the samples are all modulated by the annealing atmosphere. Notably, the samples annealed in air (GMO-air) exhibit a saturation magnetization as strong as 170% times that of the samples annealed in pure O$_2$ gas (GMO-O$_2$), which can be quantitatively explained by oxygen vacancy (V$_O$) controlled ferromagnetism due to bound magnetic polarons established between delocalized hydrogenic electrons of V$_O$s and local magnetic moments of Mn$^{2+}$, Mn$^{3+}$, and Mn$^{4+}$ ions in the samples. Our results provide insights into mechanism-based tuning of RTFM in Ga$_2$O$_3$ and may be useful for design, fabrication, and application of related spintronic materials.

## 1. INTRODUCTION

Owing to the co-existence of charge and spin degrees of freedom in the diluted magnetic semiconductors (DMSs), tremendous attention has been aroused in semiconductor spintronics.[1-4] As a promising wide-bandgap material that has been studied for a long time for various applications including solar-blind detectors and high-frequency power devices,[5,6] very recently gallium oxide ($Ga_2O_3$) has attracted much interest for its potentials in spintronic applications.[7-10] It is found that the ferromagnetism (FM) in $Ga_2O_3$ can be realized by doping with various metallic elements, such as Fe[11], Ni[12], Cr[13], Sn[14] and Mn[15-22]. Among them, Mn is the most investigated dopant in $Ga_2O_3$. For instance, Hayashi *et al.* observed room-temperature ferromagnetism (RTFM) in Mn-doped γ-$Ga_2O_3$ which was considered to be related to holes in the mid-gap Mn band.[15] Pei *et al.* reported that the RTFM in Mn-doped β-$Ga_2O_3$ could be explained by carrier-mediated double exchange model.[16] However, Guo *et al.* found that highly resistive $Ga_2O_3$ could not provide enough mobile carriers to induce the FM in double exchange model, and the FM in Mn-doped β-$Ga_2O_3$ should come from the coupling between Mn ions and oxygen vacancies ($V_O$s) since both saturation magnetization (*Ms*) and coercive field ($H_C$) increased with the concentration of Mn dopants ($Mn^{2+}$ and $Mn^{3+}$ ions).[17] The calculations based on density functional theory suggested that the RTFM of Mn-doped β-$Ga_2O_3$ should come from the strong *p-d* coupling and the delocalization of O-*2p* orbitals.[18] Very recently, Peng *et al.* reported that both FM and antiferromagnetism (AFM) existed in the $Mn^+$ ion-implanted β-$Ga_2O_3$, where FM was related to the coupling of Ga vacancies ($V_{Ga}$s) in the high resistance region and AFM came from MnO phases in the samples.[19] Obviously, although much work has been done on ferromagnetic Mn-doped $Ga_2O_3$, the mechanism underlying the ferromagnetism remains controversial. Therefore, for both basic science and potential applications, it is important and necessary to further investigate the mechanism by modulating the concentrations of vacancy defects and various Mn ions in different valence states and their possible couplings in $Ga_2O_3$.

In this work, Mn-doped β-$Ga_2O_3$ (GMO) pre-films are first prepared by polymer-assisted deposition (PAD) and then annealed in air atmosphere and pure $O_2$ gas to obtain GMO-air and GMO-$O_2$ polycrystalline thin films, respectively. The structures and concentrations of various Mn in different valence states ($Mn^{2+}$, $Mn^{3+}$ and $Mn^{4+}$ ions) and vacancy defects of the obtained films are characterized and the effects of annealing atmosphere on the magnetic, optical, and electrical properties of the films are then investigated.

## 2. EXPERIMENTAL DETALS

### 2.1 Specimen preparation

The polyethyleneimine (PEI) ((M.W. 10,000, 99%, Sigma-Aldrich) and Ethylene Diamine Tetraacetic Acid (EDTA) (99.9%, Sigma-Aldrich) are added to deionized water in a certain proportion to prepare the polymer solution. Subsequently, gallium nitrate hydrate ($GaN_3O_9·xH_2O$, 99.9%, Sigma-Aldrich) is added to the polymer solution to obtain Ga-polymers solution, and manganese chloride tetrahydrate ($MnCl_2·4H_2O$, 99.9%, Sigma-Aldrich) is added to the polymer solution to obtain Mn-polymers solution. Then the unbound ions need to be filtered out from the Ga-polymers solution and the Mn-polymers solution by Amicon cells. The Mn-doped Ga precursor (GaMn-polymers) solution is prepared by adding the Mn-polymers solution to the Ga-polymers solution, which is then dropped on single crystalline *c*-plane sapphire substrate to prepare pre-films using 30s-spin-coating at 3000 rpm. The GaMn-polymers are tested by thermalgravimetric analysis and differential scanning calorimeter (TGA-DSC) (TA instruments, Q40) to study the changes of weight and calorie during annealing in air and pure $O_2$ gas. Figure 1(a) shows the TG and DSC curves of the GaMn-polymers in air. The exothermic peaks at 290°C and 484°C represent the decompositions of the GaMn-polymers, and the endothermic peak at 588°C represents the crystallization of Mn-doped gallium oxide. The TG and DSC curves in pure $O_2$ gas show similar results. Based on the TG-DSC results, the pre-films are heated up to 550 °C for 1 h to completely decompose PEI and EDTA, and then annealed in air and pure $O_2$ gas at 750°C to obtain GMO-air and GMO-$O_2$ films, respectively. The annealing in oxygen-rich conditions is necessary for obtaining crystalline samples.[23] The large-scale (~2cm×2cm) GMO-air and GMO-$O_2$ films are translucent brown, as shown in in Figure 1(b). By changing the concentration of the precursor solution and the number of suspension coatings, dozens to hundreds of nanometer-thick GMO films are obtained. For comparison, undoped β-$Ga_2O_3$ films are also prepared.

**2.2 Characterization**

The morphologies of the GMO films are depicted by atomic force microscope (AFM, Benyuan, CSPM5500). X-ray diffraction (XRD, Bruker D8, Empyrean analytical diffractometer) using a Cu Kα radiation source and high-resolution transmission electron microscopy (HRTEM, Titan Themes Cubed G2 300) are used to characterize crystalline structures. Cross-sectional samples for TEM are prepared by focus ion beam (FIB) milling in a Helios system (Tescan, LYRA 3 XMU). Elemental composition and valence are examined by X-ray photoelectron spectroscopy (XPS, Kratos AXIS Supra). Optical properties are evaluated using UV-visible spectrophotometer (PerkinElmer Lambda 950 UV-Vis-NIR) and photoluminescence (PL, LS-45). Magnetic and electrical properties are measured by Superconducting Quantum Interference Devices (SQUID, Quantum Design, MPMS-XL-5) and four-probe resistivity tester (FPPRT, ST-2722), respectively.

## 3. RESULTS AND DISCUSSION

### 3.1 Structural Properties

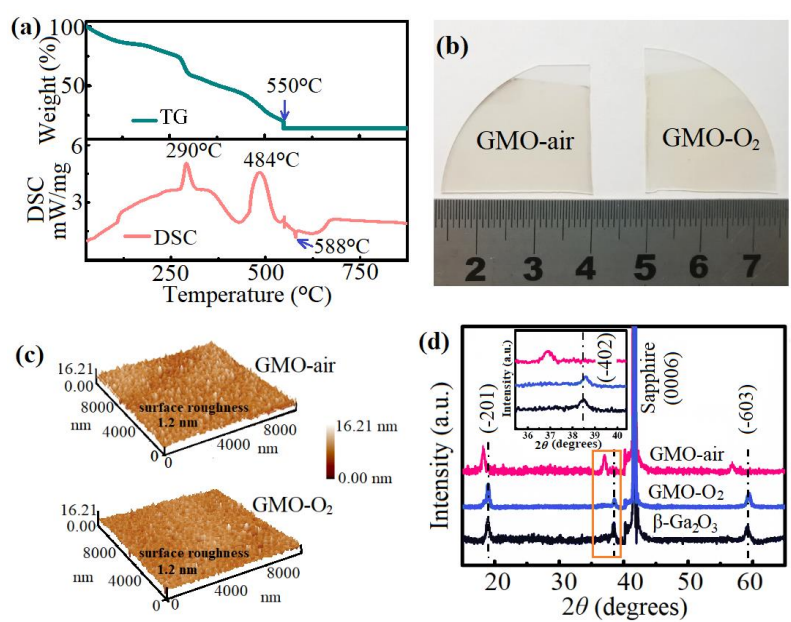

**FIGURE 1** (a) TG and DSC curves of the GaMn-polymers in the air. (b) The digital photos of the GMO films grown on the sapphire substrates. (c) 3D AFM images of the GMO films grown on the sapphire substrates. (d) XRD patterns of the PAD-grown

GMO-air, GMO-O$_2$ and undoped β-Ga$_2$O$_3$ films on the *c*-plane sapphire. The inset is the enlarged view of XRD patterns around 38°.

The three-dimensional atomic force microscope images in Figure 1(c) indicate that the surface roughness of the GMO films is ~ 1.2 nm. The crystal structures of the obtained films are studied by XRD. As shown in Figure 1(d), three diffraction peaks belong to (-201), (-402), and (-603) groups of β-Ga$_2$O$_3$ are observed in the samples, indicating the (-201) preferential growth orientation of the films.[22] The GMO-air and GMO-O$_2$ films are composed of poly crystals with an average diameter around 20.7 nm and 31.2 nm, respectively, estimated by the Scherrer formula. Importantly, no peaks of Mn-related secondary phases (clusters, oxides or compounds) are observed, because the metallic elements (Ga and Mn) are uniformly mixed in the polymers during the PAD process[23, 24] and Mn-related secondary phases are difficult to form in β-Ga$_2$O$_3$.[17, 20] The observation of single β phase of Ga$_2$O$_3$ is consistent with the single crystallization peak in the DSC curve of the GaMn-polymers in Figure 1(a). The inset is the enlarged view of $\theta$-$2\theta$ XRD patterns around 38°. Interestingly, comparing with those of un-doped β-Ga$_2$O$_3$, the diffraction peaks of GMO-O$_2$ are just slightly right-shifted, while the diffraction peaks of GMO-air are dramatically left-shifted. The (-402) inter-planar spacing values of un-doped, GMO-O$_2$ and GMO-air can be calculated using the Bragg equation and obtained as 0.232 nm, 0.230 nm and 0.240 nm, respectively. Since the size order of the cation radii in the GMO films are as follows: Mn$^{2+}$ (0.83 Å)> Mn$^{3+}$ (0.64 Å)> Ga$^{3+}$ (0.62 Å)> Mn$^{4+}$ (0.53 Å),[17] the slightly smaller inter-planar spacing of GMO-$O_2$ and the much larger inter-planar spacing of GMO-air indicate that more $Mn^{4+}$ ions are incorporated in GMO-$O_2$ and more $Mn^{2+}$ ions are incorporated in GMO-air, which will be quantitatively confirmed by the HRTEM and XPS results later.

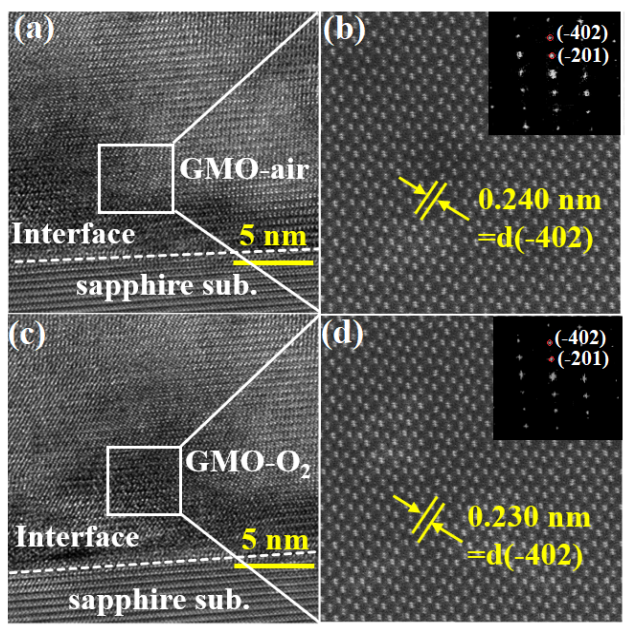

**FIGURE 2** (a) Cross-sectional HRTEM image of GMO-air film. (b) Spherical aberration TEM image of GMO-air film. (c) Cross-sectional high-resolution TEM image of GMO-$O_2$ film. (d) Spherical aberration TEM image of GMO-$O_2$ film. The insets show the corresponding FFT.

The microstructures of the GMO films are further investigated by HRTEM. The Cross-sectional HRTEM images of the GMO-air and GMO-$O_2$ films are shown in Figures 2(a) and 2(c), respectively. The corresponding spherical aberration HRTEM images of the GMO-air and GMO-$O_2$ films are shown in Figures 2(b) and 2(d), respectively. Both the GMO-air and GMO-$O_2$ films are highly crystalline and grown along the direction of (-201), consistent with the XRD results. Clear crystalline structures can be seen and no Mn-related clusters or precipitates are observed in both cases. The inter-planar spacings along (-402) azimuth of the GMO-air and GMO-$O_2$ films are about 0.240 nm and 0.230 nm, respectively, consistent with the XRD results.

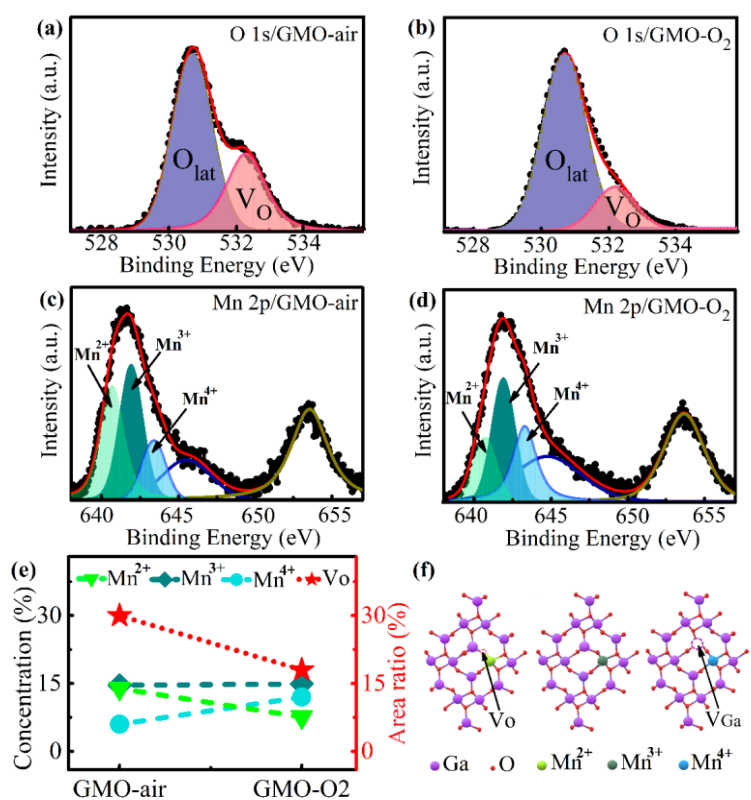

**FIGURE 3** O 1s peaks of (a) GMO-air and (b) GMO-$O_2$. Mn 2p peaks of (c) GMO-air and (d) GMO-$O_2$. (e) The doping concentrations of different Mn ions and the area ratio of $V_{Os}$-related peak in the GMO films. (f) The schematic diagram of formation of $Mn^{2+}$, $Mn^{3+}$ and $Mn^{4+}$ ions in $Ga_2O_3$.

The chemical states of the elements in the GMO films are studies by XPS. By fitting the XPS whole scanning spectra (not shown), the atomic ratios between oxygen and cation of GMO-air and GMO-$O_2$ films are obtained as 1.42 and 1.45, respectively. The stoichiometric ratio is less than 1.5 in $Ga_2O_3$, indicating the existence of vacancy defects.

As shown in Figures 3(a) and (b), each O 1s peak of GMO-air and GMO-O$_2$ could be fitted into two sub-peaks located at 530.7 eV and 532.1 eV, corresponding to O ions surrounded by metal atoms (O$_{lat}$) and V$_{Os}$ in the metal bonding matrix, respectively.[24,25] The area ratio of V$_{Os}$-related peak of GMO-air and GMO-O$_2$ are estimated to be 30% and 18%, respectively, indicating that there are more V$_{Os}$ in GMO-air. The Mn doping concentrations in GMO-air and GMO-O$_2$ both are about 34.5%, estimated by fitting the XPS whole scanning spectra. Figures 3(c) and (d) show the Mn 2p spectra of the GMO-air and GMO-O$_2$ films, respectively. The main peak can be fitted into three sub-peaks located at about 640.6 eV, 641.8 eV and 643.2 eV, corresponding to Mn$^{2+}$, Mn$^{3+}$ and Mn$^{4+}$ ions, respectively.[17, 26] The concentrations of Mn$^{2+}$, Mn$^{3+}$ and Mn$^{4+}$ ions in GMO-air are obtained as 13.9%, 14.6% and 6.0%, respectively, while the concentrations of Mn$^{2+}$, Mn$^{3+}$ and Mn$^{4+}$ ions in GMO-O$_2$ are 7.6%, 14.9% and 12.0%, respectively. The doping concentrations of different Mn ions and the area ratio of V$_{Os}$-related peak are shown in Figure 3(f). Much more Mn$^{2+}$ ions in GMO-air are responsible for the increase of inter-planar spacing shown in Figure 1(d). Meanwhile, more Mn$^{4+}$ ions in GMO-O$_2$ are responsible for the decrease of inter-planar spacing shown in Figure 1(d). In general, substitutional Mn dopants in Ga$_2$O$_3$ mainly exist as Mn$^{3+}$ ions when there are no other defects around.[18] When there is a V$_O$ nearby a Mn$^{3+}$ ion, two extra electrons are introduced in the β-Ga$_2$O$_3$ lattice, one of which enters the Mn d-shell and covert the Mn$^{3+}$ ion into an Mn$^{2+}$ ion.[18,26] When there is a V$_{Ga}$ nearby a Mn$^{3+}$ ion, the V$_{Ga}$ acts as an acceptor and captures a 3$d$ electron from the Mn d-shell, converting the Mn$^{3+}$ ion into an Mn$^{4+}$ ion.[18,26] The schematic of formation of Mn$^{2+}$, Mn$^{3+}$ and Mn$^{4+}$ ions in the GMO films is shown in Figure 3(e). It is worthwhile noting that this is the first time that $Mn^{4+}$ ions are observed in Mn-doped $Ga_2O_3$. Based on the assumption that one $Mn^{4+}$ ion corresponds to one $V_{Ga}$, the higher concentration of $Mn^{4+}$ ions in GMO-$O_2$ means higher $V_{Ga}$ concentration in GMO-$O_2$, which is reasonable since pure $O_2$ provide more oxygen than air.

## 3.2 Optical and Ferromagnetic Properties

Figure 4(a) shows the optical transmittance measured by UV-visible spectrophotometer of the GMO films. The GMO-air film exhibits slightly lower optical transmittance than that of the GMO-$O_2$ film, consistent with the results observed in Figure 1(b). The inset is the $(\alpha h\nu)^2$ versus $h\nu$ plot of the GMO films. By fitting the plot of $(\alpha h\nu)^2$ versus $h\nu$, the band gaps of GMO-air and GMO-$O_2$ are obtained as 4.98 eV and 5.18 eV, respectively. It is more $V_{OS}$ in GMO-air that results in lower optical transmittance and smaller band gap, owing to the contributions of more occupied defect levels.[27,28] Figure 4(b) shows the PL spectra of the GMO films. The emission band can be divided into four bands centered at about 345 nm, 415 nm, 440 nm and 480 nm, respectively. The ultraviolet emission band located at about 345 nm corresponds to the recombination of self-trapped excitons, which is an intrinsic process.[29] The emission peaks centered at 415 nm, 440 nm (violet light) and 480 nm (blue light) are originated from the electron-hole recombination formed by $V_{OS}$, or from the recombination of Ga-O vacancy pair.[30,31] The PL intensity at the violet and blue light regions of the GMO-air films are stronger than that of GMO-$O_2$ films, which can be attributed to more defects such as $V_{OS}$ in the GMO-air films. The electrical measurements show that the resistivity values of the GMO-air and GMO-O$_2$ films are 1.2×10$^4$ Ω·cm and 9.1 ×10$^4$ Ω·cm, respectively, and the majority carriers are electrons (n-type), which can be roughly explained by the fact that more V$_O$s exist in GMO-air films since V$_O$s act as donor-type defects to provide electrons in Ga$_2$O$_3$. In short, the tuning of the optical and electrical properties is mainly resulted from the large difference between the V$_O$ concentrations in the GMO-air and GMO-O$_2$ films.

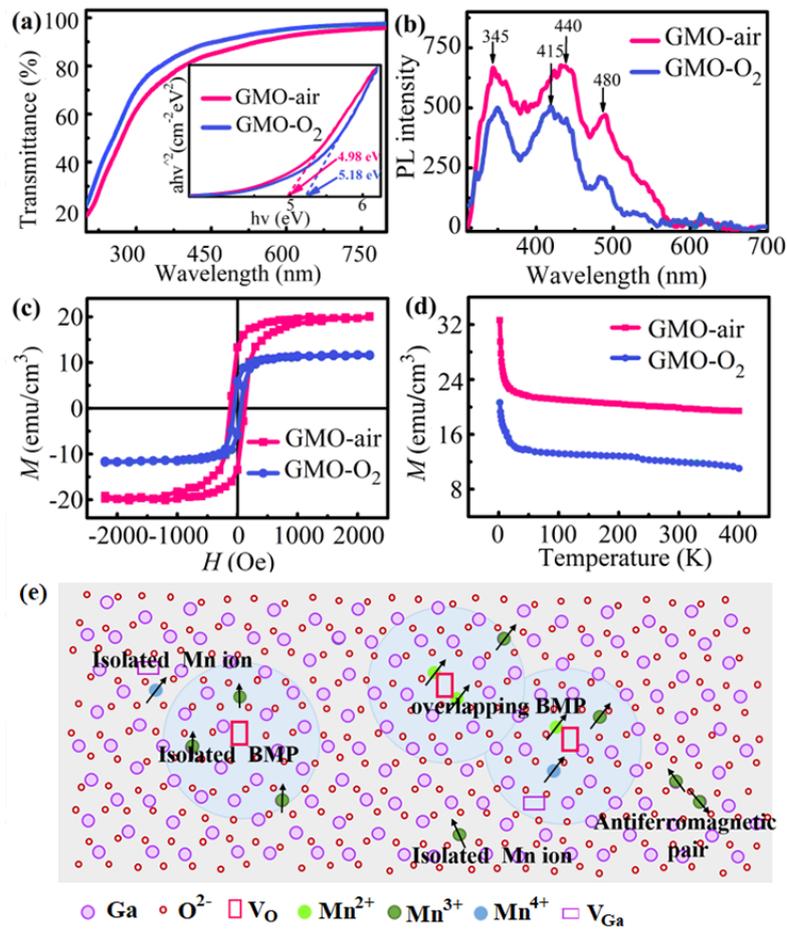

**FIGURE 4** (a) Optical transmittance spectra of the GMO films. Inset is the $(\alpha h\nu)^2$ versus $h\nu$ plot. (b) Room-temperature PL spectra of the GMO films excited at 250 nm. (c) M-H curves of the GMO films at 300 K. (d) Temperature-dependent magnetization of the GMO films. (e) Schematic of bound magnetic polarons in the GMO films.

Finally, the magnetic properties of the GMO samples are investigated. Figure 4(c) shows the magnetic field dependent magnetization (*M-H*) curves of the GMO films obtained at 300 K. During the *M-H* measurements, *H* is applied in plane which is parallel to the easy magnetic axis of Mn-doped β-Ga$_2$O$_3$ films preferentially grown along the (-201) orientation.[20] Both GMO films display typical hysteresis loops, which are clear signals of ferromagnetism. The $M_S$ values of the GMO-air and GMO-O$_2$ films are 19.7 emu/cm$^3$ and 11.6 emu/cm$^3$, respectively. The $H_C$ values of the GMO-air and GMO-O$_2$ films are 120 Oe and 58 Oe, respectively. Obviously, GMO-air exhibits stronger $M_S$ and $H_C$ values, 1.7 and 2.1 times those of GMO-O$_2$, respectively. The temperature-dependent magnetization (*M-T*) curves of the GMO-air and GMO-O$_2$ films measured in field-cooling mode under 2000 Oe are shown in Figure 4(d), indicating ferromagnetism up to temperatures higher than 400 K. The magnetizations of both samples increase very slowly from 400 K down to 20 K, and then increase steeply at temperatures lower than 20 K, which could be ascribed to the presence of paramagnetic components inducing additional magnetization at low temperatures.[15] Notably, the *M-T* curves also reveal much stronger magnetization in GMO-air films than that in GMO-O$_2$ films.

To understand the difference between the ferromagnetic properties in the GMO-air and GMO-O$_2$ films, we should point out that the ferromagnetism comes from bound magnetic polarons (BMPs)[32] which are closely related to the V$_O$ defects in the GMO films. In the β-Ga$_2$O$_3$ lattice, a V$_O$ acts as a donor and provides two additional electrons confined in hydrogenic orbitals.[33] When the donor concentration exceeds a critical value, the hydrogenic orbitals overlap to form impurity bands, making the localized hydrogenic electrons become delocalized. The overlap between the delocalized electrons and local spin moments of Mn ions within the hydrogenic orbitals leads to ferromagnetic exchange coupling between them to form BMPs, [33] as shown Figure 4(e). When the amount of BMPs is sufficient, percolation occurs and ferromagnetic ordering is spontaneously induced. It is noted that Mn ions not included in hydrogenic orbitals will not participate in ferromagnetic exchange coupling. In addition, $V_{Ga}$s may produce weak local spin moments, but it is difficult for them to form long-range magnetic ordering.[10] In short, the ferromagnetism mainly depends on the concentrations of $V_O$s and Mn ions that are coupled with each other. According to Hund's rule, the spins contributed by an $Mn^{2+}$, an $Mn^{3+}$ and an $Mn^{4+}$ ion are 4/2, 5/2 and 4/2, respectively.[33,34] In other words, an $Mn^{2+}$ and an $Mn^{4+}$ ion provide the same local spin moment, while an $Mn^{3+}$ ion provides more local spin moment than the other two. Interestingly, GMO-air and GMO-$O_2$ have approximately the same concentration of $Mn^{3+}$ ions and same concentration of total Mn dopants. This means that, if the $V_O$ concentration was high enough so that all the Mn ions were included in hydrogenic orbitals, GMO-air and GMO-$O_2$ would exhibit the same magnetization, which obviously is not the case in our work. In fact, the $V_O$ concentration is much lower than the concentration (~34.5%) of all the Mn ions in the GMO films. In this situation, the magnetization is determined by the $V_O$ concentration but not the concentration of the Mn ions, and the much stronger magnetization in GMO-air can only be ascribed to much more $V_O$s in it. Notably, the $V_O$ concentration in GMO-air is 1.7 times that in GMO-$O_2$, quantitatively consistent with the ratio (1.7) of the $M_s$ of GMO-air over that of GMO-O$_2$. Furthermore, the greater $H_C$ in GMO-air is also ascribed to more $V_O$s, since $H_C$ increases with the increase of pinning sites originated from $V_O$s in oxide films.[35, 36]

## 4. CONCLUSIONS

The Mn-doped β-Ga$_2$O$_3$ films with RTFM are prepared by PAD and the effects of annealing atmosphere on their structures and physical properties are studied. Our results show that the RTFM comes from $V_O$-controlled BMPs established between sufficient delocalized electrons from $V_O$s and local spin moments of Mn various ions, and dramatical (1.7 times) tuning of the ferromagnetism can be realized based on the ferromagnetism mechanism by annealing atmosphere. Our results provide insights into the defect-controlled ferromagnetism in β-Ga$_2$O$_3$ and may be useful for related electronic and spintronic applications.

## ACKNOWLEDGMENT

We are grateful to the National Natural Science Foundation of China (Grant No. 52172272). We thank the Analytical & Testing Center of Sichuan University for the XRD and XPS measurements.

**List of Figure Captions**

**FIGURE 1** (a) TG and DSC curves of the GaMn-polymers in the air. (b) The digital photos of the GMO films grown on the sapphire substrates. (c) 3D AFM images of the GMO films grown on the sapphire substrates. (d) XRD patterns of the PAD-grown GMO-air, GMO-O$_2$ and undoped β-Ga$_2$O$_3$ films on the *c*-plane sapphire. The inset is the enlarged view of XRD patterns around 38°.

**FIGURE 2** (a) Cross-sectional HRTEM image of GMO-air film. (b) Spherical aberration TEM image of GMO-air film. (c) Cross-sectional high-resolution TEM image of GMO-$O_2$ film. (d) Spherical aberration TEM image of GMO-$O_2$ film. The insets show the corresponding FFT.

**FIGURE 3** O 1s peaks of (a) GMO-air and (b) GMO-$O_2$. Mn 2p peaks of (c) GMO-air and (d) GMO-$O_2$. (e) The doping concentrations of different Mn ions and the area ratio of $V_{Os}$-related peak in the GMO films. (f) The schematic diagram of formation of $Mn^{2+}$, $Mn^{3+}$ and $Mn^{4+}$ ions in $Ga_2O_3$.

**FIGURE 4** (a) Optical transmittance spectra of the GMO films. Inset is the $(\alpha h\nu)^2$ versus $h\nu$ plot. (b) Room-temperature PL spectra of the GMO films excited at 250 nm. (c) M-H curves of the GMO films at 300 K. (d) Temperature-dependent magnetization of the GMO films. (e) Schematic of bound magnetic polarons in the GMO films.

**Author contributions**